# New pressure-induced monoclinic β-Sb$_2$Te$_3$ phase with sevenfold symmetry

Sergio Michielon de Souza[*], Daniela Menegon Trichês[*] and Claudio Michel Poffo

*Departamento de Engenharia Mecânica, Universidade Federal de Santa Catarina, Campus Universitário Trindade, S/N, C.P. 476, 88040-900 Florianópolis, Santa Catarina, Brazil*

João Cardoso de Lima and Tarciso Antonio Grandi

*Departamento de Física, Universidade Federal de Santa Catarina, Campus Universitário Trindade, S/N, C.P. 476, 88040-900 Florianópolis, Santa Catarina, Brazil*

Alain Polian and Michel Gauthier

*Physique des Milieux Denses, IMPMC, CNRS-UMR 7590, Université Pierre et Marie Curie-Paris 6, B115, 4 Place Jussieu, 75252 Paris Cedex 05, France*

**ABSTRACT**

A nanometric Sb$_2$Te$_3$ rhombohedral phase was produced from Sb and Te by mechanical alloying for 3 hours and its structural stability was studied by synchrotron X-ray diffraction (XRD) and Raman spectroscopy (RS) measurements as a function of pressure. A phase transformation from the ambient pressure rhombohedral phase into a β-Sb$_2$Te$_3$ monoclinic structure between 9.8 and 13.2 GPa is observed by XRD. This phase transformation is confirmed by the Raman spectroscopy measurements. The pressure dependence of the volume fited to a Birch-Murnaghan equation of state gives a bulk modulus $B_0 = 40.6 \pm 1.5$ GPa and $B'_0 = 5.1 \pm 0.6$. The bulk modulus of the nano-Sb$_2$Te$_3$ seems to be slightly smaller than that for its bulk counterpart (44.8 GPa).

PACS index: 61.05.cp, 61.50.Ks, 62.50.-p, 61.46.Hk

# I. INTRODUCTION

Compounds based on Bi, Sb and Te ($V_2$-$VI_3$ semiconductors) are the best known materials for thermoelectric applications at room temperature or below. $Sb_2Te_3$ is one of these compounds that has a great potential for technological applications [1-4]. Its figure of merit ZT exhibits a maximum close to room temperature.

$Sb_2Te_3$ is a layered compound that at ambient conditions crystallizes in a rhombohedral structure $R\bar{3}m$ (S.G. 166, Z = 3). One layer, i.e., one unit bonded through iono-covalent bonds, consists of five alternating sheets of Sb and Te. The succession is [Te(2)-Sb-Te(1)-Sb-Te(2)], where Sb and Te(2) atoms occupy the 6c Wyckoff sites (in the hexagonal setting) and Te(1) atoms the 3*a* sites [5]. The hexagonal *c* axis is perpendicular to the plane of the layers and the layers are bonded by van der Waals bonds. The elementary piece of this structure is a *SbTe₆* octahedron centered on a Sb atom. From this point of view, the layers are formed by two planes of adjacent edge sharing $SbTe_6$ octahedra with common Te(1) atoms.

In spite of its technological importance, the effect of high pressure on the $Sb_2Te_3$ compound has not been widely studied. Using X-ray diffraction (XRD) and electrical resistance measurements, Sakai et al. [6] observed a phase transformation at 10.7 GPa, while Jacobsen et al [7] using XRD measurements observed a phase transformation for pressures between 7 and 10 GPa. In Ref. [7] the new phase is indexed to an orthorhombic phase (I222, S.G.23), but this result is controversial.

The mechanical alloying (MA) is an efficient technique to synthesize many unique materials, such as nanometric and amorphous alloys as well as metastable solid solutions [8]. It has many advantages including low temperature processing, easy composition control, inexpensive equipment and the possibility of industrial production. Its main disadvantage is the contamination by the milling media and/or the milling atmosphere. Different techniques, including MA, have been used to produce nanometric phases [8]. A good review of the MA technique is found in Ref. 9, while the physical mechanisms involved are discussed in Refs. 10-13.

Nanometric materials have two components: crystallites of nanometric dimensions (2–100 nm) with the same structure as their crystalline counterparts, and an interfacial component, which comprises the several types of defects (grain boundaries, interphase boundaries, dislocations, etc.). Nanometric materials are metastable [8]. Thermoelectric $Sb_2Te_3$ has been prepared by several techniques, including MA [14].

There are several studies describing the effect of high-pressure on nanometric materials [15-20]. However, there are no results for $Sb_2Te_3$ in nanometric form. In this work, we report the effect of high pressure on a nanometric $Sb_2Te_3$ rhombohedral phase prepared by MA. The structural and vibrational changes with increasing pressure were studied by XRD and Raman spectroscopy (RS).

## II. EXPERIMENTAL PROCEDURE

A binary Sb-Te mixture of high-purity elemental powders of Sb (Aldrich 99.999%) and Te (Alfa Aesar 99.999%, -100 meshes) in the proportion 2:3 atomic was sealed together with several steel balls of 11.0 mm in diameter into a cylindrical steel vial under argon atmosphere. The ball-to-powder weight ratio was 7:1. The vial was mounted on a SPEX Mixer/mill, model 8000. The temperature was kept close to the ambient temperature by a ventilation system. After 3 hours of milling, the measured XRD pattern was indexed to the $Sb_2Te_3$ rhombohedral stable phase and milling process was interrupted.

A membrane diamond anvil cell (DAC) [21] with an opening that allowed probing up to 28° of 2θ was used. A small amount of $Sb_2Te_3$ powder was compacted between diamonds to a final thickness of approximately 15 µm. A small chip of this preparation, about 80 µm in diameter, was then loaded into a stainless-steel gasket hole of 150 µm diameter. Neon gas was used as a pressure-transmitting medium because (i) it is one of the softests materials, (ii) it is chemically inert, and (iii) it has no luminescence and no Raman activity. The pressure was determined through the fluorescence shift of a ruby sphere [22] loaded in the sample chamber. The quasi-hydrostatic conditions were controled throughout the experiments by monitoring the separation and widths of $R_1$ and $R_2$ lines. *In situ* XRD patterns as a function of pressure were acquired at the XRD1 station of the ELETTRA synchrotron radiation facility. This diffraction beamline is designed to provide a monochromatized, high-flux, tunable x-ray source in the spectral range from 4 to 25 keV [23]. The present study was performed using a wavelength of 0.068881 nm. The detector was a 345-mm imaging plate from MarResearch. The sample-to-detector distance was calibrated by diffraction data from Si powder loaded in the diamond anvil cell. The data were collected with a 10 min exposure time. The two-dimensional diffraction patterns were converted to intensity versus 2θ using the fit2D software [24] and analyzed by the Rietveld method using the GSAS package [25].

For the Raman measurements as a function of pressure, one particle of approximately 50 x 60 x 20 µm$^2$ was loaded in the DAC. The Raman spectra and ruby luminescence were recorded in the backscattering geometry by means of a Jobin-Yvon T64000 Raman triple spectrometer and a liquid-nitrogen-cooled charge coupled device multichannel detector. An excitation line of λ = 514.5 nm of an Ar laser was used for excitation and focused down to 5 µm with a power of about 20 mW at the entrance of the DAC. The acquisition time was 1800 s. The Raman frequencies were determined from a fit of the peaks to a Lorentzian profile. The frequency accuracy was better than 1 cm$^{-1}$.

**III. RESULTS AND DISCUSSION**

The as-milled powder has a microstructure that consists of a $Sb_2Te_3$ rhombohedral matrix with Te particles. This powder, in form of a pellet, was sealed in a quartz tube evacuated at about 10$^{-3}$ Torr and annealed at 583 K for 9 h, followed by cooling in air. As-milled and annealed powders were characterized through XRD, RS, differential scanning calorimetry (DSC) and photoacoustic absorption spectroscopy (PAS) measurements. The results were reported in Ref. 26, and will be not repeated here, but the main difference between them is that the annealed powder properties are very similar to that of the bulk samples.

*A. High pressure XRD measurements*

*In situ* XRD measurements on as-milled nanometric $Sb_2Te_3$ rhombohedral (phase I) powder were performed at increasing pressure up to 19.2 GPa. Fig. 1 shows some representative XRD patterns. Up to 9.8 GPa, the XRD patterns correspond to the rhombohedral phase I. With increasing pressure, the peaks are shifted toward higher *2θ* values, their intensity decreases and they broaden. At 9.8 GPa, a shoulder at about *2θ* = 13.8$^o$ is observed in the XRD pattern, indicating the nucleation and growth of a new phase (phase II). This new phase is completely formed at 13.2 GPa and remains up to below 15.5 GPa where it disappears almost completely. Between 15.2 and 19.2 GPa, despite the presence of diffraction peaks from the crystalline neon and the metallic gasket preventing the correct interpretation of the XRD patterns, two new peaks located between $2\theta = 13.8°$ and $18°$ indicate the emergence of a new $Sb_2Te_3$ phase (phase III).

$A_2B_3$ (A = Bi, and Sb) and (B = Te and Se) form an isomorphous family of compounds. In an unpublished study, we investigated the effect of high-pressure up to 31.7 GPa on the mechanically alloyed $Bi_2Te_3$ rhombohedral powder. Besides the ambient rhombohedral phase (phase I), at least three new high-pressure phases (phases II, III and IV) were observed. To date, only the results concerning the effect of high pressure on the phase I were published [27]. Comparison between the XRD patterns for the phase II of $Sb_2Te_3$ and $Bi_2Te_3$ shows a resemblance (Fig. 2).

Einaga et al. [28] and Zhu et al. [29] investigated the effect of high pressure on the bulk $Bi_2Te_3$ rhombohedral phase. The new high pressure phases are similar to those observed by us for nanometric $Bi_2Te_3$. Zhu et al. [29] reproduced the XRD pattern measured for the phase II and III assuming seven- (C2/m, S.G. 12) β-$Bi_2Te_3$ and eightfold (C2/c, S.G. 15) γ-$Bi_2Te_3$ monoclinic structures, while Einaga et al. [28] reproduced the XRD pattern measured for the phase IV assuming a structural model analogous to a substitutional Bi-Te binary alloy (60 atomic % tellurium), where the Bi and Te atoms are distributed in the bcc lattice sites ($Im\bar{3}m$, S.G. 229).

The XRD patterns for the nanometric $Sb_2Te_3$ rhombohedral phase (phase I) at various pressures were refined using the Rietveld method [30]. The lattice parameters *a* and *c*, the *c/a* ratio, and the volume V as a function of pressure are shown and compared with the literature data of Ref. 6 and 7 in Figs. 3 (a), (b) and (c), respectively. At low pressures, the *c* parameter decreases faster than *a* as demonstrated by the initial slope of the *c/a* ratio versus pressure. This anisotropic compressibility is often observed in low dimensionality compounds [31, 32], where atoms in adjacent layers are linked through van der Waals forces. Above ~4 GPa, the *c/a* ratio slope changes sign, in qualitative agreement with the results reported in Refs. 6 for $Sb_2Te_3$ and 27 and 33 for $Bi_2Te_3$. This is an indication that the repulsive part of the Van der Waals bonds starts to play an important role. The volume as a function of pressure *V*(p) obtained from the Rietveld refinement was fitted to a Birch-Murnaghan equation of state (BM EOS) [34]:

$$p = \frac{3}{2} B_0 X^5 (X^2 - 1) \left\{ 1 + \frac{3}{4} (B' - 4)(X^2 - 1) \right\} \qquad (1)$$

where $X = (V_0/V)^{1/3}$. A fit of the experimental *V*(p) in the stability range of rhombohedral phase to a BM EOS (Fig. 3 (c)) gives $B_0 = 40.6 \pm 1.5$ GPa and $B'_0 = 5.1 \pm 0.6$. The literature reports a value of $B_0 = 44.823$ GPa for bulk $Sb_2Te_3$ compound [35]. Jacobsen et al. [7] reported $B_0 = 60.8$ GPa with $B'_0 = 3.4$ using a Vinet EOS.

Jacobsen et al. [7] also reported the existence of an electronic topological transition (ETT) or Lifschitz transition [36] around 3 GPa. Such a transition is due a modification of the Fermi surface topology due to hydrostatic and non-hydrostatic compression and has been shown to influence strongly the thermoelectrical properties of compounds [37], particularly in the case of $Bi_2Te_3$ [38]. The insufficient number of data points in the XRD measurements prevents us to evidence the ETT in $Sb_2Te_3$.

## B. β-$Sb_2Te_3$ (phase II)

The structural model proposed by Zhu *et al.* [29] for the sevenfold β-$Bi_2Te_3$ monoclinic (C2/m S.G. 12) was used as initial data to simulate the measured XRD pattern at 13.2 GPa (phase II) of as-milled $Sb_2Te_3$ powder. The best simulation was achieved for the structural parameters listed in Table I. Fig. 4 shows the excellent agreement between the experimental and simulated patterns.

There are constraints that the proposed sevenfold β-$Sb_2Te_3$ monoclinic structure must satisfy: i) the density must be larger than or equal to that of the lower pressure structure, and ii) the smallest Sb-Te interatomic distance must not be smaller than roughly the sum of the atomic radii of Sb and Te atoms. If the covalent atomic radii of Sb (0.141 nm) and Te (0.137 nm) [39] are considered, the smallest Sb-Te interatomic distances should not be less than approximately 0.278 nm. At 9.8 GPa, from the Rietveld simulation a density value of 7.639 g/cm$^3$ was obtained for the rhombohedral phase (phase I) while, after the transition, a value of 8.224 g/cm$^3$ was obtained for the sevenfold monoclinic phase (phase II). This value is 8% larger than before phase transition. Using the structural data listed in Table I in the Crystal Office 98 software [40], the smallest calculated Sb-Te interatomic distance is 0.275 nm, which is compatible with the estimated Sb-Te interatomic distances.

*C. Raman spectroscopy under pressure*

The $Sb_2Te_3$ rhombohedral phase crystallizes in $R\bar{3}m$ symmetry, and the normal modes at the Γ point of the Brillouin zone are classified according to the irreducible representations of this point group [41]

$$\Gamma = 2(A_{1g} + E_g) + 3(A_{2u} + E_u) \qquad (2)$$

Because of the inversion symmetry, there is exclusion between the Raman and infrared activity, and the *g* modes are Raman active whereas the *u* modes are IR active (one $A_u$ and one $E_u$ are acoustic modes). Richter *et al*. [42] measured the Raman spectra at ambient conditions (except the low frequency $E_g^1$) and more recently Sosso et al. [41] calculated the frequency shift of the Raman and IR active modes at ambient conditions for the rhombohedral phase. The calculated [41] (experimental [42]) wave number of the Raman active modes are: $\sigma(E_g^1) = 46$ cm$^{-1}$ (-), $\sigma(E_g^2) = 113$ cm$^{-1}$ (112 cm$^{-1}$), $\sigma(A_{1g}^1) = 69$ cm$^{-1}$ (69 cm$^{-1}$) and $\sigma(A_{1g}^2) = 166$ cm$^{-1}$ (165 cm$^{-1}$). The IR modes are listed in Ref. 41, where each Raman irreducible representation is correlated with a set of displacement pattern in the *a–b* plane (E modes) and along the *c*-axis (A modes).

Figure 5 shows the measured RS spectra for the nanometric Sb$_2$Te$_3$ powder at several pressures. The RS spectrum measured at ambient pressure agrees quite well with that measured out of the DAC and shown in Ref. 26. The as-milled powder has a microstructure formed by a Sb$_2$Te$_3$ rhombohedral matrix and Te particles [26], and their Raman active modes are marked in Fig. 5. At ambient pressure, the wave number of the Te Raman active modes are $A_1 = 122$ cm$^{-1}$ and $E = 141.3$ cm$^{-1}$ (E). The effect of high pressure on the structure and on the Raman active modes of trigonal Te was investigated by Partharasathy and Holzapfel [43] and Richter et al [44], respectively. All the Te Raman active modes decrease with increasing pressure [44] as shown by the red solid lines (online) in Fig. 5 and the first structural phase transition occurs at about 4 GPa [43]. For the nanometric Sb$_2$Te$_3$ rhombohedral powder, all the Raman active modes originating from phase I become weaker, broader, shift to higher wave numbers with increasing pressure up to 10.6 GPa, and disappear completely at 13.8 GPa. The behavior of these modes suggests a gradual transformation of nanometric Sb$_2$Te$_3$ phase I into the β-Sb$_2$Te$_3$ phase II, which is completed at 13.8 GPa.

The sevenfold β-Sb$_2$Te$_3$ monoclinic structure (C2/m) has 6 Raman active modes at the Γ point of the Brillouin zone, which are given by the irreducible representation [45]

$$\Gamma = 4A_g + 2B_g. \tag{3}$$

Figure 6 shows the Raman spectrum measured for the sevenfold β-Sb$_2$Te$_3$ monoclinic phase for 13.8 and 25.8 GPa, where one can see that between 50 and 650 cm$^{-1}$ there are three broad bands labeled A, B and C in Fig. 7, which could be followed up to the maximum pressure reached in this experiment, although a second phase transformation was observed by XRD around 15.8 GPa (see *supra*). Nevertheless, the intensity of the two highest frequency modes decrease with increasing pressure, which may be an indication of a metallization of the compound.

In phase I, the pressure dependence of the wave number may be approximated with a standard second order polynomial:

$$\omega(P) = \omega_0 + AP + BP^2, \tag{4}$$

where $\omega_0$ is the wave number in cm$^{-1}$ at zero pressure and P in GPa. The wave number of the Raman active modes ω were obtained through a fitting procedure using Lorentzian profiles. The obtained dependences are:

$$\omega(P)A_{1g}^1 = 70.58 + 4.42P - 0.13P^2$$

$$\omega(P)A_{1g}^2 = 166.35 + 2.75P - 0.04P^2. \tag{5}$$

$$\omega(P)E_g^2 = 114.20 + 2.86P - 0.88P^2$$

Figure 7 shows this pressure dependence.

The effect of high pressure on the Raman active modes can be better understood by considering the derivative of Eq. (4): $d\omega/dP = A + 2BP$

Figure 8 shows the derivative of analytical expressions (5) obtained from fits for $E_g^2$, $A_{1g}^1$ and $A_{1g}^2$ modes. From this figure one can see that with increasing pressure, the derivative of $A_{1g}^1$ mode varies faster than that of the $E_g^2$ and $A_{1g}^2$ modes. This is not surprising considering the displacement pattern of these different modes [41,44]. Indeed, the $A_{1g}^1$ mode is a "respiratory" mode of the layer parallel to the *c*-axis, and hence, the rapidly varying interlayer van der Waals interaction is one of the main restoring forces for this mode. On the contrary, for the other two vibrations, this interaction is only implied at the second order. This behavior is very similar to that of the vibrational modes of GaS [46]. From another point of view, $E_g^2$ and $A_{1g}^2$ imply the same interatomic interactions although the displacement is along the *c*-axis for the $A_{1g}$ mode and in the layer plane for the $E_g$ one, and hence similar pressure dependence is not surprising. The $A_{1g}^2$ mode is slightly less sensitive to pressure increase of than the $E_g^2$ mode. Above 9 GPa, the effect of pressure is stronger in the $A_{1g}^1$ mode than in the $E_g^2$ and $A_{1g}^2$ modes, as seen in Fig. 5.

The Grüneisen parameter $\gamma_0$ describes the effect of high pressure on the volume of structural lattice of the material, and consequently, on the phonons vibrations. The zero-pressure mode Grüneisen parameters $\gamma_0$ are determined using the equation [47, 48]

$$\gamma_0 = \frac{B_0}{\omega_0}\left(\frac{\partial \omega}{\partial P}\right)_{P=0} \tag{6}$$

where $B_0$ and $\omega_0$ are the bulk modulus in GPa and wave number in cm$^{-1}$ at zero pressure. By XRD, we obtained (see preceding paragraph) $B_0 = 40.6 \pm 1.5$ GPa. This value is ≈ 9% smaller than that reported in the literature for bulk rhombohedral Sb$_2$Te$_3$ [35]. Using $B_0$ and $\omega_0$ values in Eq. (6), the $\gamma_0$ values for the $A_{1g}^1$, $E_g^2$ and $A_{1g}^2$ modes are 2.55 (2.82), 0.67 (0.74), and 1.04 (1.15), respectively. The numbers between parentheses were calculated using $B_0 = 44.823$ GPa. Again, one observe here that the pressure affect much more the $A_{1g}^1$ mode than the other.

**IV. CONCLUSIONS**

Nanometric $Sb_2Te_3$ rhombohedral powder was produced by MA. Its structural and vibrational properties were investigated through *in situ* high-pressure XRD and RS measurements. With increasing pressure, several structural transformations were seen, but due to the diffraction peaks associated with the crystallization of neon gas (used as a pressure transmitting medium) and those associated to the gasket, only the data of phase I and phase II could be used. The XRD pattern measured at 13.2 GPa (phase II) for the $Sb_2Te_3$ was well reproduced assuming a sevenfold β-$Bi_2Te_3$ monoclinic structure (C2/m). From the analysis of RS spectra measured with increasing pressure for the as-milled nanometric $Sb_2Te_3$ powder, the pressure dependence of the Raman active modes were established. The weak van der Waals interlayer interaction translates more on the $A_{1g}^1$ mode variation than in that of the $E_g^2$ and $A_{1g}^2$ ones. This is because this interaction is more involved in the displacement pattern of the former than in that of the later.

**ACKNOWLEDGMENTS**

This study was part of the PhD thesis of one of the authors (S.M.S) and it was financially supported by the Brazilian-French CAPES/COFECUB Program (Project No. 559/7). We thank ELETTRA synchrotron (Trieste, Italy) for the XRD measurements as a function of pressure. We are indebted to Jean-Claude Chervin, Pascal Munch and Gilles Le Marchand for technical support.

* Present adress: Departamento de Física, Universidade Federal do Amazonas, 3000 Japiim, 69077-000 Manaus, Amazonas, Brazil

**FIGURES**

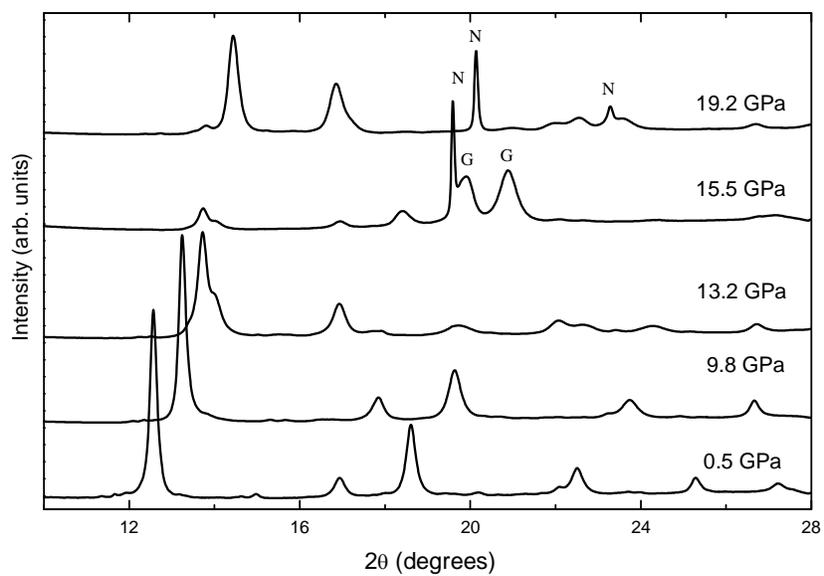

Figure 1: XRD patterns measured with increasing pressure for the nanometric $Sb_2Te_3$ rhombohedral powder. The highest pressure reached was 19.2 GPa.

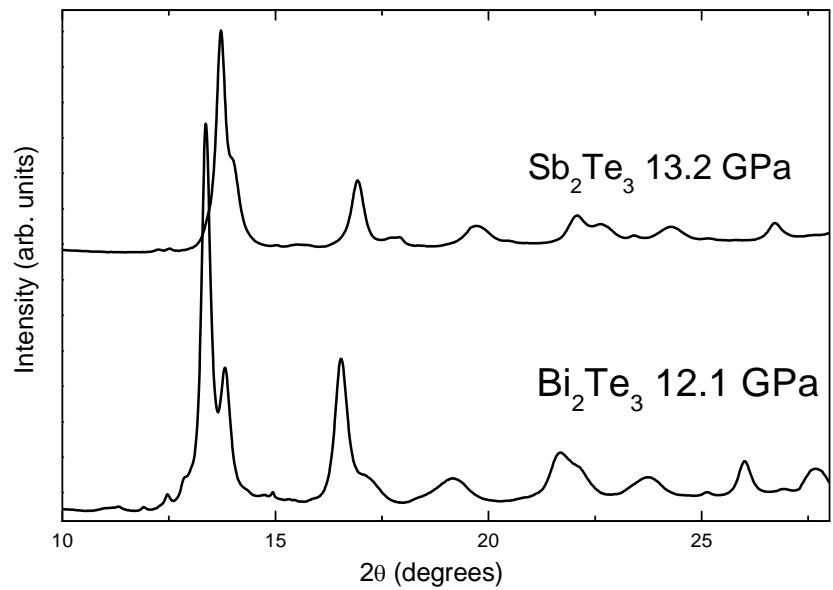

Figure 2: Comparison between the XRD patterns measured for phase II in nanometric Bi$_2$Te$_3$ and Sb$_2$Te$_3$ rhombohedral powders at 12.1 and 13.2 GPa, respectively.

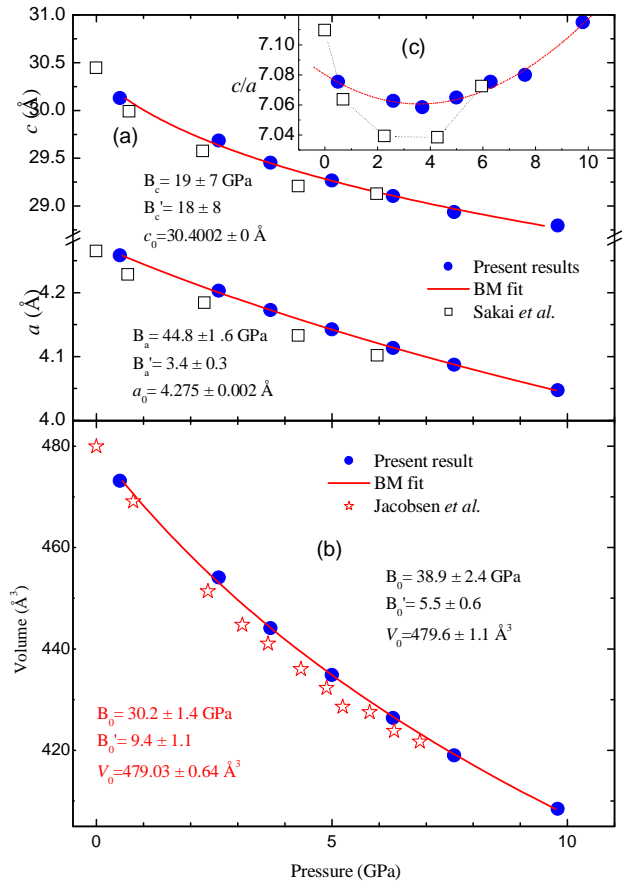

Figure 3 (color online): Pressure dependence of the structural parameters of $Sb_2Te_3$ deduced from Rietveld refinements compared with published results: (A) Lattice parameters; (B) c/a ratio. (C) Volume. Full circles: present results; open squares: Ref. 6; open stars: Ref. 7. The solid lines are the fits to a Birch-Murnaghan equation of states; dotted and dashed lines are guides for the eyes.

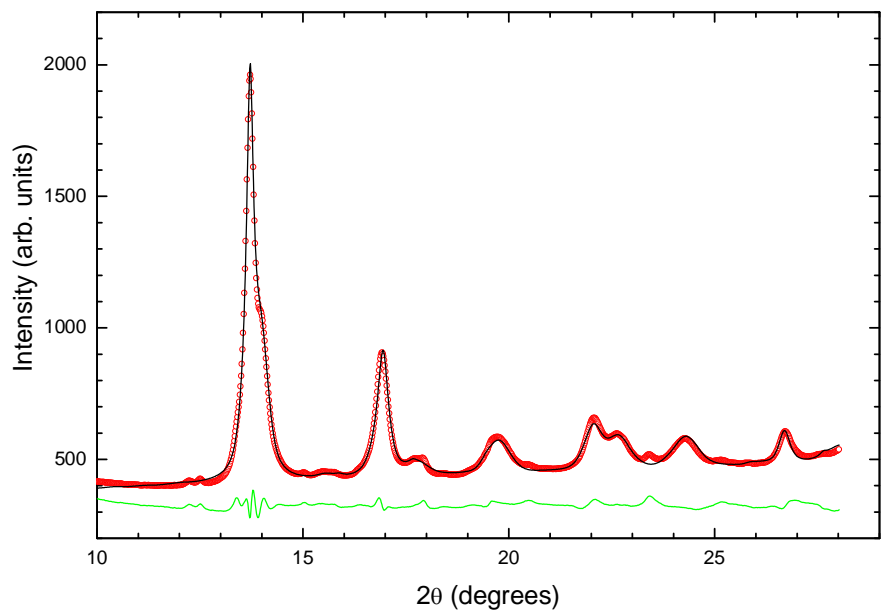

Figure 4: XRD pattern of high-pressure phase II of $Sb_2Te_3$ at 13.2 GPa (open circles). Solid line represents the Rietveld simulation for the structural data listed in Table I. The bottom line is the residual intensities.

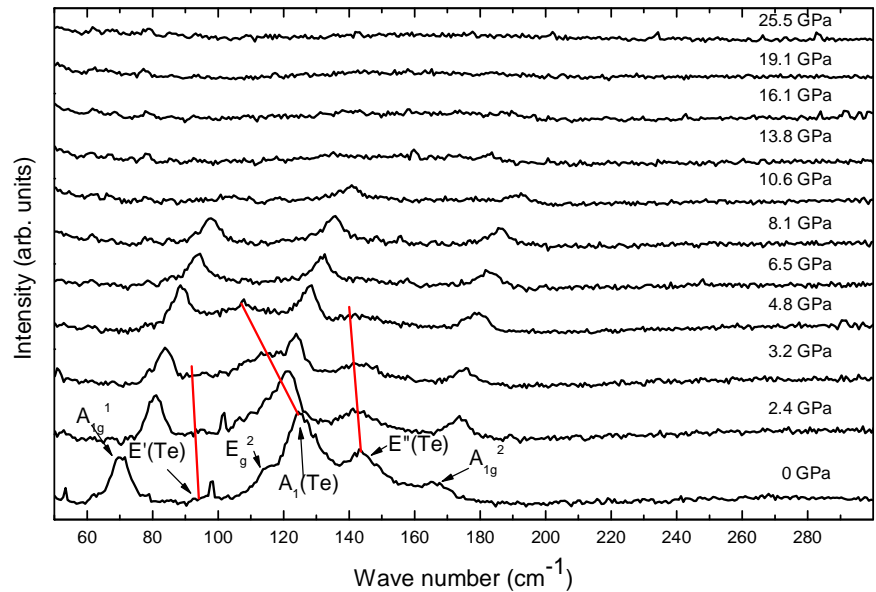

Figure 5 (color online): Raman spectra measured with increasing pressure for the nanometric $Sb_2Te_3$ rhombohedral powder. The excitation wavelength was $\lambda = 514.5$ nm and highest pressure was 25.5 GPa. The red full lines represent Raman peaks from tellurium impurities (see text).

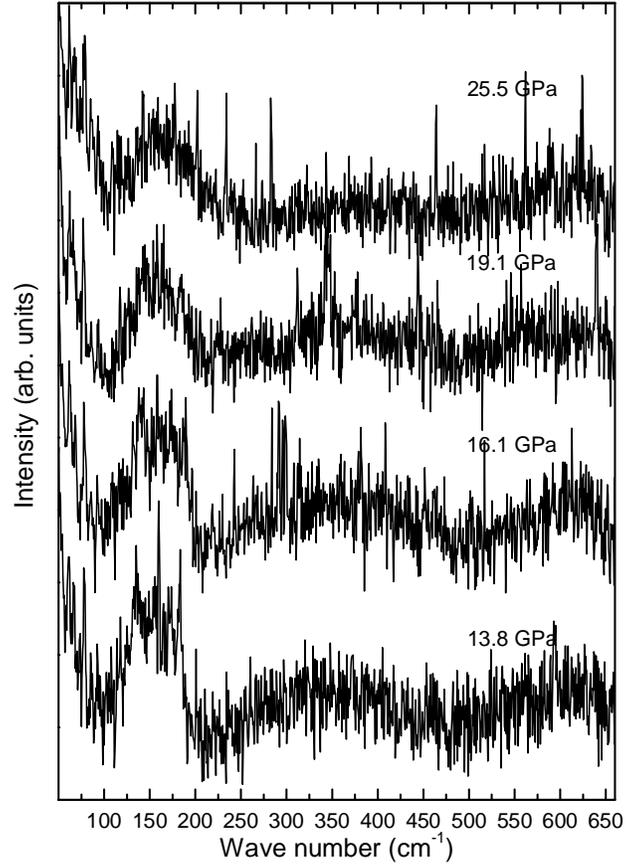

Figure 6: Raman spectra measured for the sevenfold β-$Sb_2Te_3$ monoclinic phase between 13.8 and 25.5 GPa, The excitation wavelength was λ = 514.5 nm and highest pressure was 25.5 GPa.

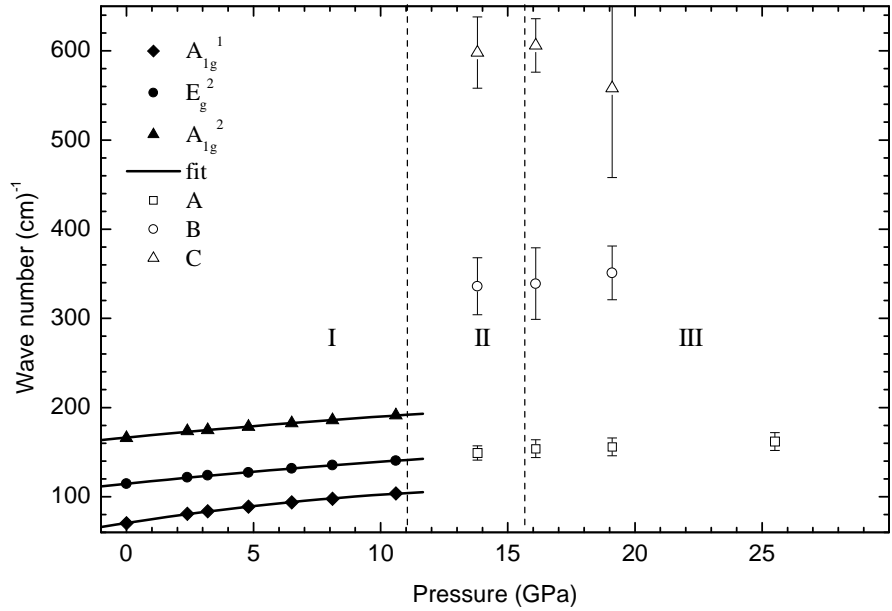

Figure 7: Pressure dependence of the Raman active modes of $Sb_2Te_3$ up to 25.5 GPa through two phase transitions. Symbols represent the experimental data and solid lines in phase I the polynomial fits (see Eq (4)) in the text. The vertical dashed lines are at the positions of the phase transitions, and the latin numbers are for the phases. Error bars on the modes of the high pressure phases are estimations.

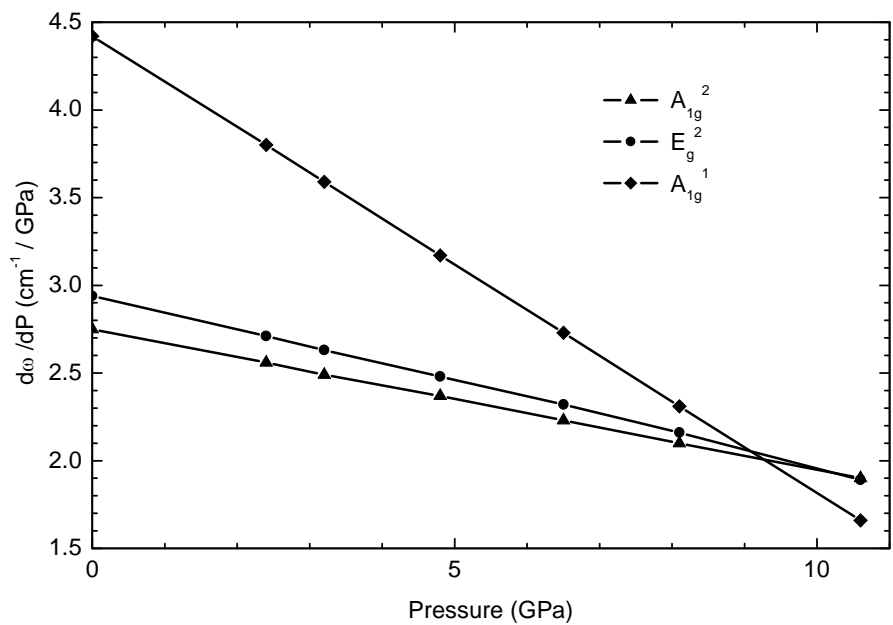

Figure 8: Derivatives of analytical expressions (5) representing the wave number values measured with increasing pressure. The symbols represent the experimental data and the solid lines are guides for the eyes.

Table I: Structural data obtained from the Rietveld simulation for the sevenfold β-$Sb_2Te_3$ monoclinic structure (phase II).

| Atoms | x | y | z | Lattice parameters (Å) and angle (°) |
|---|---|---|---|---|
| Sb1 | 0.1971 | 0 | 0.2079 | $a$ = 14.3717 |
| Sb2 | 0.4599 | 0 | 0.2340 | $b$ = 4.0138 |
| Te1 | 0.2321 | 0 | 0.3965 | $c$ = 17.0901 |
| Te2 | 0.0451 | 0 | 0.6106 | $β$ = 149.130° |
| Te3 | 0.3470 | 0 | 0.9853 | |